\documentclass[aps,prd,10pt,twocolumn,nofootinbib]{revtex4}

\usepackage{epsfig}
\usepackage{bm}
\usepackage{latexsym}
\usepackage{natbib}
\usepackage{url}
\usepackage{dcolumn}
\usepackage{color}
\usepackage{amsfonts,amssymb,amsmath}
\usepackage{graphicx,epsfig}
\usepackage{psfrag}
\usepackage{subfigure}
\usepackage{hyperref}
\hypersetup{colorlinks=true}

\newcommand{\ba}{\begin{array}}
\newcommand{\ea}{\end{array}}
\newcommand{\be}{\begin{equation}}
\newcommand{\ee}{\end{equation}}
\newcommand{\bea}{\begin{eqnarray}}
\newcommand{\eea}{\end{eqnarray}}

\begin{document}

\title{Extranatural Inflation Redux}

\author{Mansi Dhuria$^{a,}$\footnote{mansidh@phy.iitb.ac.in}}
\author{Gaurav Goswami$^{b,}$\footnote{gaurav.goswami@ahduni.edu.in}}
\author{Jayanti Prasad$^{c,}$\footnote{jayanti@iucaa.in}}
\affiliation{
$^a$ Department of Physics, Indian Institute of Technology Bombay, Mumbai 400076, India \\
$^b$ School of Engineering and Applied Science, Ahmedabad University, Ahmedabad 380009, India \\
$^c$ The Inter-University Centre for Astronomy and Astrophysics, Pune 411 007, India}

\begin{abstract}
The success of a given inflationary model crucially depends upon two features: its predictions for  observables such as those of the Cosmic Microwave background (CMB) and its insensitivity to the unknown ultraviolet (UV) physics such as quantum gravitational effects. 
Extranatural inflation is a well motivated scenario which is insensitive to UV physics by construction. 
In this five dimensional model, the fifth dimension is compactified on a circle and the zero mode of the fifth component of a bulk $U(1)$ gauge field acts as the inflaton.
In this work, we study simple variations of the minimal extranatural inflation model in order to improve its CMB predictions while retaining its numerous merits. 
We find that it is possible to obtain CMB predictions identical to those of e.g. ${\cal R} + {\cal R}^2$ Starobinsky model of inflation and show that this can be done in the most minimal way by having two additional light fermionic species in the bulk, with the same $U(1)$ charges. We then find the constraints that CMB observations impose on the parameters of the model.
\end{abstract}

\maketitle

\section{Introduction}

In the Standard Model of cosmology, one posits, among other things, that the early Universe was incredibly homogeneous, isotropic and spatially flat but still had small, adiabatic and Gaussian density perturbations which had a nearly scale invariant power spectrum. In the recent past, a successful combination of theory and experiments has validated this picture. 
Cosmic inflation \cite{Starobinsky:1980te,Kazanas:1980tx,Guth:1980zm,Linde:1981mu,Albrecht:1982wi,Starobinsky:1979ty,Hawking:1982cz,Starobinsky:1982ee,Guth:1982ec} is a mechanism by which the microscopic theories (very similar to the ones we routinely use to explain the behaviour of elementary particles in e.g. colliders) can give rise to a Universe which would look very much like the one posited in the Standard Model of Cosmology. 
One of the questions worth addressing in near future is whether one can learn more about the details of inflation.

In the next decade, the upcoming cosmological experiments 
\cite{Finelli:2016cyd,Andre:2013afa,Abazajian:2016yjj,Creminelli:2015oda,Munoz:2016owz} will help us better understand this era in the early history of the Universe.
Though better observational data will surely help in this quest, a careful look at the literature suggests that realistic inflation model building is a formidable task irrespective of the available data \cite{Baumann:2014nda}. E.g. small field inflationary potentials suffer from overshooting problem while large field inflation can be extremely sensitive to the unknown UV physics. Thus, even though there are many models of inflation which seem to give CMB predictions which agree with recent observations (such as those of Planck experiment \cite{Ade:2015lrj}), a model of inflation which does not suffer from other theoretical problems and still agrees with experiments is a rarity. In this paper, we wish to look for a model of inflation which achieves this.

A well known example of a model in which there are supposedly no issues of UV sensitivity, at least at the level of field theory, is extranatural inflation \cite{ArkaniHamed:2003wu} 
\footnote{It is well known that extranatural inflation requires a rather small (${\cal O} (10^{-3})$) value of a 4D gauge coupling and this is challenging to achieve in known UV completions such as string theory 
\cite{Banks:2003sx,ArkaniHamed:2006dz,delaFuente:2014aca}. We do not address UV completion of extranatural inflation in this work.
}.
As we explain in \textsection{\ref{ENI}}, there are many reasons because of which this scenario is preferred over natural inflation \cite{Freese:1990rb,Adams:1992bn}.
In extranatural inflation, the zero mode of the fifth component of a bulk Abelian gauge field acts as the inflaton. This field is clearly a scalar under four dimensional coordinate transformations. The potential of the inflaton is generated by loop corrections of light fermions (charged under the bulk gauge group) present in the bulk. 
However, it turns out that this minimal scenario gives CMB predictions identical to those of natural inflation which is increasingly getting disfavoured with newer CMB data. The most recent CMB data however is completely consistent with the famous ${\cal R} + {\cal R}^2$ model of inflation of Starobinsky \cite{Starobinsky:1980te}.
It may thus be worthwhile looking for variations of the minimal scenario of extranatural inflation which give predictions identical to those of  Starobinsky model.
We propose to achieve this by adding extra fermionic species in the bulk. The observational data then constrains the charge of these fermions under the bulk gauge group. 

It turns out that the effect of adding such extra fermions in the bulk is the addition of more sinusoidal functions to the potential of natural inflation.
Though scenarios with such potentials have been studied for a long time in the context of both natural inflation \cite{Kim:2004rp,Higaki:2014pja,Higaki:2014sja,Czerny:2014wza,Choi:2014rja,Kobayashi:2016vcx,Kappl:2014lra,Kappl:2015esy} and extranatural inflation \cite{Feng:2003mk,Wan:2014fra,Croon:2014dma,GonzalezFelipe:2007uy,Kohri:2014rja}, the possibility of obtaining CMB predictions similar to those of Starobinsky model has not been explored (see however \cite{Higaki:2015kta,Li:2015taa}).

This paper is organized as follows: in \textsection \ref{ENI}, we begin by explaining the issue of UV sensitivity of inflation and then review the relevant details of extranatural inflation.
Then, in \textsection \ref{CMB}, we explain our scenario and its possible CMB predictions, show that adding a single extra fermion in the bulk does not sufficiently improve the CMB predictions of extra-natural inflation and then present the minimal variation of extranatural inflation whose CMB predictions are identical to those of Starobinsky model. We then find the values of all the parameters and scales in the model using the constraints obtained from CMB observations. Finally, we conclude in \textsection \ref{discussion}.

{\bf Notation:} We work with $\hbar = c = 1$ units, moreover, $M_p$ is 4D Planck mass, $M_{\rm Pl}$ is the 4D reduced Planck mass,
$M_p^{(5)}$ is 5D Planck mass, $\ell_P$ is the 4D Planck length,
$R$ is the radius of extra dimension, $L = 2\pi R$ and $\cal R$ is the Ricci scalar.

\section{Extranatural inflation} \label{ENI}

Before getting into the details of extra-natural inflation, we revisit the issue of UV sensitivity of inflation, particularly large field inflation \cite{Baumann:2014nda}.

\subsection{UV sensitivity of inflation}

Even though during inflation, the energy density of the inflaton field dominates the Universe, the inflaton would be but one field in a Lagrangian which would, in any realistic picture, contain many fields. In fact, there would at least be Standard Model fields in the same Lagrangian. Moreover, we might need more fields to explain e.g. neutrino masses, Dark Matter or Baryon Asymmetry of the Universe or to solve the cosmological constant problem. Moreover, there must be new degrees of freedom which would show up near Planck scale which would help unitarize the graviton-graviton scattering at Planck scale. 

Every theory is to be interpreted as a Wilsonian EFT with a physical cut-off. The Wilsonian effective action can be obtained from the UV theory by integrating out the physics above a UV cut-off $\Lambda_0$.
E.g., if, in the path integral of the theory, one integrates out all the fields except the inflaton and also integrates out all the high frequency modes (above some scale $\Lambda_0$) of the inflaton, one would obtain the Wilsonian effective Lagrangian of the inflaton which would be of the form \cite{Srednicki:2007qs}
 \begin{eqnarray} \label{wilson}
 {\cal L}_{\rm eff} [\phi] =  {\cal L}_{\ell} [\phi] + \sum_{i=1}^{\infty} 
 c_i \frac{\phi^{4+2i}}{\Lambda_0^{2i}} + d_i \frac{(\partial \phi)^2 \phi^{2i}}{\Lambda_0^{2i}} + \nonumber \\
 e_i \frac{(\partial \phi)^{2(i+1)} }{\Lambda_0^{4i}} + \cdots  
 \; ,
 \end{eqnarray}
where, it is assumed that a $Z_2$ symmetry holds good in the UV theory.

For most observables, when one performs experiments at energies well below $\Lambda_0$, the higher dimension operators have negligible effect. 
But this is not always true: e.g. mass of an elementary scalar is highly sensitive to {\it all} the higher dimension operators (see \cite{Srednicki:2007qs} for details). 
E.g. the $m^2 \phi^2$ inflation happens to be such that all the Wilson coefficients in the above Lagrangian except the $m^2$ term, vanish. The question is what ensures that this will happen? 
In the context of inflation, this problem is closely related to the so called eta problem:  given that the higher dimension operators could renormalize $m^2$ (with typical contributions of the order of $\Lambda_0^2$), how come $m^2 \ll H^2$?

{\it Rolling beyond the cut-off:} The Wilson effective action is valid only when one performs experiments at energies below the cut-off scale. If the Lagrangian contains a higher dimension operator, at high enough energies, unitarity is violated.
On the other hand, during large field inflation, the field rolls by a super-Planckian amount. 
Notice that if we have a large field inflation $\phi > M_p$ which means $\phi > \Lambda_0$, this means that the contribution of higher powers in Eq (\ref{wilson}) is even higher (unless the coefficients somehow compensate for this).
This raises the question: when the field rolls by an amount greater than the cut-off, is the Wilson action even valid? It is often argued that since energy density during inflation is conveniently sub-Planckian, there is no problem if the field vev changes by super-Planckian values; but, in the Wilson action, since field excursion is super-Planckian, unless all the infinite Wilson coefficients are all guaranteed to be small, we would surely have a problem. 

Finally, one may be concerned how the inflaton potential may get affected by unknown UV physics e.g. loop corrections due to heavy particles, the effects of virtual black holes or gravitational and other instantons in e.g. string theory
\cite{Hebecker:2016dsw,Alonso:2017avz}.

\subsection{Symmetries of UV theory: global and gauge}

Symmetries of the theory which UV completes the theory of the scalar can cure this problem. This is because symmetries kill all  the higher dimension operators which can contribute to the mass of the scalar.

E.g. if one assumes that there is a global shift symmetry in the UV theory, this will set all the $c_i, d_i$ to zero. But this will also set even $m$ and $\lambda$ to be zero. One could then generate $m$ by breaking the global shift symmetry softly by an independent sector.
 
As far as the coefficients $e_i$ are concerned, they need not be small since for a homogeneous inflating background, $(\partial \phi)^2 = {\dot \phi}^2 + {(\nabla \phi)}^2 \approx {\dot \phi}^2 = 2 \epsilon H^2$ which is suppressed by $\epsilon$, the Hubble slow-roll parameter. 
Moreover, the quantum correction to the mass of the scalar due to these derivative operators is $\delta m \propto m$ while their contributions to a scattering amplitude at energies lower than $\Lambda_0$ is anyway negligible.

The most familiar example implementing this ideas is Natural Inflation \cite{Freese:1990rb,Adams:1992bn}. To begin with, the inflaton is assumed to be the Goldstone mode of a spontaneously broken global $U(1)$ symmetry. This causes its potential to vanish at all orders in perturbation theory.
If one now also assumes that this symmetry is anomalous i.e. though it exists in the classical theory, it is broken by quantum effects, then, gauge instantons generate a potential which is a cosine at the leading order. However, the requirement of having large field slow-roll inflation causes the scale of spontaneous breaking of $U(1)$ to be super-Planckian. Since there are reasons to suspect that there can be no continuous global symmetries in quantum gravity (see \cite{Kallosh:1995hi} and the discussion in section 4 of \cite{Banks:2010zn}), one must find out alternatives to the most basic natural inflation (e.g. by having multiple $U(1)$s \cite{Kim:2004rp} or by taking into account spinodal instabilities \cite{Albrecht:2014sea}). 
In stark contrast, in extranatural inflation \cite{ArkaniHamed:2003wu}, instead of global symmetries, a gauge symmetry forbids the coefficients $c_i$ and $d_i$ so that the unknown UV physics has negligible effects on the potential of the inflaton.
A lot of recent work \cite{Banks:2003sx,ArkaniHamed:2006dz,delaFuente:2014aca} has been devoted to trying to understand the issue of possible UV insensitivity of extranatural inflation and similar models. In this work, however, we'd assume that the inflaton potential for extranatural inflation can be protected from unknown UV effects and focus on improving its CMB predictions.

\subsection{Extranatural inflation in a nutshell}
In the rest of the paper, we would restrict our attention to Quantum Field Theory in a five dimensional spacetime where the fifth dimension is compactified on a circle i.e. the spacetime in the absence of gravity is $M_4 \times S^1$. The coordinates on this 5D spacetime are denoted as $(x^{\mu},y)$.

{\it Boundary Conditions:} Since the extra-dimensional coordinate $y$ is identified to $y+2 \pi R$, for all fields $\Phi(x^\mu,y) \sim \Phi(x^\mu,y+2 \pi R)$.

By mode expansion, one can verify that a single 5D (i.e. bulk) Abelian gauge field $A_M$ is equivalent to the following fields (it is easiest to see the field content in the so-called ``almost-axial" gauge, see e.g. \cite{Sundrum:2005jf} for details):
  \begin{itemize}
   \item $A_5^{(0)}$, which is a gauge invariant, massless 4D scalar with no tree level potential (this will act as the inflaton),
   \item $A_\mu^{(0)}$, which has a residual gauge invariance, a massless 4D vector,
   \item $A_\mu^{(n)}$, an infinite tower of massive 4D vectors (the Kaluza-Klein i.e. KK modes of the vector).
  \end{itemize}

Before proceeding, it is worth noting that in five dimensions, a gauge field has mass dimension $3/2$ while the corresponding 5D gauge coupling $g_5$ has mass dimension $-1/2$. Of course the 4D gauge coupling is dimensionless, in fact
\begin{equation} \label{eq:unitarity}
g_4 = \frac{g_5}{\sqrt{2\pi R}} \; .
\end{equation}
One can define the dimensionless and gauge-invariant field 
  \begin{equation}
  \theta (x) = g_5 \oint dy A_5 (x^{\mu},y) \; ,
  \end{equation}
which is also the gauge-invariant Wilson loop of $A_5$ along the extra dimension (and, as we shall see, is going to be very simply related to the inflaton field). It is easy to verify that this integral will pick only the contributions from the zero mode of the fifth component of the bulk Abelian gauge field i.e. $A_5^{(0)}$.

If bulk matter is present, a potential for $A_5^{(0)}$ and hence the inflaton is readily generated.
If there is one bulk matter field which is charged under the gauge symmetry of the 5D Abelian gauge field (e.g. a bulk complex scalar field or a bulk spinor field), and hence has a charge $Q$, then, the gauge covariant derivative in its Lagrangian will be given by $D_M = \partial_M - i Q g_5 A_M $.
From 4D point of view, this bulk matter field will give rise to an infinite tower of KK modes. Thus, from 4D point of view, the 4D-scalar-field $A_5^{(0)}$ has coupling to all these infinite matter fields. So, every KK mode of the matter field will generate a Coleman-Weinberg potential for $A_5^{(0)}$.
For a bulk matter field with mass $m_a$ and $U(1)$ charge $Q_a$, the Coleman-Weinberg potential of $\theta$ due to the KK modes of this bulk matter field is given by 
(see \cite{Hosotani:1983xw,Delgado:1998qr,Hatanaka:1998yp} for some early references,
\cite{Antoniadis:2001cv} for a particularly accessible derivation and
\cite{Feng:2003mk,delaFuente:2014aca} for some relatively recent papers)
\begin{equation} \label{eq:pot-0}
V (\theta) = \pm \frac{3}{64 \pi^6 R^4} \left[ \sum_{n=1}^{\infty} c_n e^{-2\pi R n m_a} Re ( e^{i n Q_a \theta} ) \right] \; ,
\end{equation}
where,
\begin{equation}
c_n = \frac{1}{n^5} + \frac{2 \pi R m_a}{n^4} + \frac{(2 \pi R m_a)^2}{3 n^3} \; ,
\end{equation}
and the $+$ sign is for fermionic matter while $-$ sign is for bosonic matter.

If the bulk matter is massless (or has a mass very small as compared to $R^{-1}$), taking the $m_a \rightarrow 0$ limit in the above expression gives
  \begin{equation} \label{eq:pot}
    V (\theta) = \pm \frac{3}{64 \pi^6 R^4} \sum_{n=1}^{\infty} \frac{\cos (n Q_a \theta)}{n^5} \; .
  \end{equation}

Finally, for the sake completion, when one turns on gravity, the spacetime will have a curved geometry but will still retain the topology of $M_4 \times S^1$. The radius of the circle will then be different at different points of 4D spacetime and will be determined from the vev of a scalar field called the radius modulus (or radion). The 5D Einstein gravity gives rise to
    \begin{itemize}
     \item $h_{55}^{(0)}$, which is gauge invariant, a massless 4D scalar with no tree level potential (this is the radius modulus or the radion, its vev will have to be stabilized to a value large enough so that the inflaton potential can be kept protected from unknown UV effects),
     \item $h_{5\mu}^{(0)}$, which has a residual gauge invariance, a massless 4D vector (the gravi-photon),
     \item $h_{\mu \nu}^{(0)}$, which has a residual gauge invariance, a massless spin two particle (the familiar 4D graviton), 
     \item $h_{\mu \nu}^{(n)}$, an infinite tower of massive KK gravitons.
    \end{itemize} 

\noindent {\it The various fields in extranatural inflation:} We thus have a 5D Abelian gauge field, 5D Einstein gravity, bulk matter (and if required: bulk cosmological constant and brane tension). We would be interested in solutions in which the radion is stabilized (i.e it sits at the bottom of its potential) so that the physical size of the extra dimension is fixed. On the other hand, the inflaton is rolling down and hence the effective 4D cosmological constant is positive and dominates the dynamics of the universe: thus, the 4D universe is undergoing inflation. In this work, the vaccum energy at the minimum of the inflaton potential shall be assumed to be zero.

\subsection{Connection to natural inflation}

If we specialize to the case of the potential generated due to just light fermions in the bulk and notice that the subsequent terms in Eq (\ref{eq:pot}) are suppressed so that the term with $n=1$ dominates, the potential due to only one fermion in the bulk will be of the form 
\begin{equation}
V(\theta) \approx \frac{3}{64 \pi^6 R^4} {\cos (Q_a \theta)} \; .
\end{equation}
The dimensionless field $\theta (x)$ is canonically normalized to \cite{ArkaniHamed:2003wu} 
\begin{equation}
\phi = \frac{\theta}{g_4 (2 \pi R)} \; ,
\end{equation}
which is the inflaton, this gives 
\begin{equation}
V(\phi) \approx \frac{3}{64 \pi^6 R^4} {\cos (Q_a g_4 2 \pi R \phi)} \; .
\end{equation}
If we now define
\begin{equation} \label{eq:decayconst}
 f = \frac{1}{2 \pi R g_4} \; ,
\end{equation}
then 
\begin{equation}
V(\phi) \approx \frac{3}{64 \pi^6 R^4} {\cos \left( \frac{Q_a \phi}{f} \right)} \; .
\end{equation}
If one adds an appropriate constant to this potential in order to keep the minimum of the potential at zero vacuum energy
\footnote{This is equivalent to assuming a solution to the cosmological constant problem \cite{Polchinski:2006gy}.}, 
one obtains the potential of natural inflation
\begin{equation} \label{nat-infl}
V = \Lambda^4 \left[ 1 + \cos \left( \frac{\phi}{f_{\rm eff}} \right) \right] \; ,
\end{equation}
where, the overall factor $\Lambda$ in Eq (\ref{nat-infl}) is given by 
\begin{equation}\label{lambda}
\Lambda^4 = \frac{3}{64 \pi^6 R^4} \; .
\end{equation}
It is clear from Eq (\ref{nat-infl}) that CMB data is sensitive to the ``decay constant" $f_{\rm eff} = f/Q_a$. Thus, it is only the ratio of $f$ (defined by Eq (\ref{eq:decayconst})) and $Q_a$ which can be determined from the data.

In summary, if we have just one light fermion in the bulk with $U(1)$ charge unity, then, the predictions of extra-natural inflation are identical to those of natural inflation to a very good accuracy. It is however noteworthy that the most recent CMB data disfavours natural inflation at $2 \sigma$ statistical significance \cite{Ade:2015lrj}.

\subsection{Merits of extranatural inflation}

Before proceeding, it is worth noting that in extranatural inflation   
even though the effective scale $f$ appears to be super-Planckian in 4D description, there is no super-Planckian mass scale involved in 5D description. This is because a super-Planckian ``axion decay constant" can be obtained
by having a small 4D gauge coupling
\begin{equation}
\frac{f}{M_{\rm Pl}} = \frac{1}{2\pi g_4 (R M_{\rm Pl})} \; .
\end{equation}
Moreover, heavy particles which are uncharged under the bulk $U(1)$ gauge symmetry can not affect the inflaton potential while although the potential gets affected by the loops of heavy particles which are charged under the bulk gauge symmetry, this effect is exponentially suppressed (see Eq (\ref{eq:pot-0})).

The only remaining concern is the super-Planckian excursion of the inflaton since there is a possibility that quantum gravitational effects could still affect the potential. In \cite{ArkaniHamed:2003wu}, it is mentioned that since the super-Planckian decay constant originates from sub-Planckian mass scales, they expect that quantum gravitational effects on the potential go at most as $\sim e^{-2\pi R M_5}$ (the exponential is suppressed by the Euclidean action of a relativistic particle going around the extra dimension).
A look at the recent literature lends credence to the notion that the jury is still out on the validity of such estimates \cite{Hebecker:2016dsw,Alonso:2017avz}.
We also note that successful large field inflation in extranatural inflation is achieved by assuming the 4D gauge coupling to be too small. This may be harmless in field theory but, as is well known \cite{Banks:2003sx}, strongly resists any embedding in string theory. In this work, we shall not be exploring these fascinating issues any further but instead turn to observational constraints.

\section{CMB observations and parameters} \label{CMB}

We saw in the last section that one light fermion in the bulk leads to a potential which, to leading order, is of the form of a cosine. To aid the discussion, we would use the phrase ``first fermion" to refer to the bulk fermion whose loop corrections generate the potential of natural inflation.
We now turn our attention to variations of extra natural inflation in which additional fermions shall be present in the bulk \cite{Feng:2003mk,Bai:2014coa}. It is worth noting that we only consider the additional fermions to be light as compared to the KK scale.

Let us suppose we have one light fermion with charge $Q_a$ and then ${\cal N}$ copies of another light fermion with charge $Q$, then the potential of the inflaton would be
\begin{eqnarray} \label{eq:pot1}
 V(\phi) &=& \frac{3}{64 \pi^6 R^4} \Biggr\{ \sum_{n=1}^{\infty} \frac{1}{n^5} \cos \left( \frac{n Q_a \phi}{f} \right) \nonumber\\
 &&+ {\cal N} \sum_{n=1}^{\infty}  \frac{1}{n^5} \cos \left( \frac{n Q \phi}{f} \right) + C \Biggr\} \; .
\end{eqnarray}
The constant $C$ in the above potential is chosen such that the vacuum energy of the minimum of the potential is zero.
Since it is only the ratios $f/Q_a$ and $f/Q$ which determine the arguments of the cosines, we could set $Q_a$ to 1 and hence rescale $Q$ and $f$ accordingly.
Thus, if we have one light fermion with charge $+1$ and then ${\cal N}$ copies of another light fermion with charge $Q$, then the potential of the inflaton would be
\begin{eqnarray} \label{eq:pot2}
 V(\phi) &=& \frac{3}{64 \pi^6 R^4} \Biggr\{ \sum_{n=1}^{\infty} \frac{1}{n^5} \cos \left( \frac{n \phi}{f} \right) \nonumber\\
 &&+ {\cal N} \sum_{n=1}^{\infty}  \frac{1}{n^5} \cos \left( \frac{n Q \phi}{f} \right) + C \Biggr\} \; .
\end{eqnarray}

Notice that this is quite different from the potentials dealt with in e.g. multinatural inflation where the amplitudes, frequencies and phases of the two cosines could all be arbitrarily different from each other. 
Thus, one comes across a more constrained scenario simply due to the extra dimensional embedding of our model.
In the rest of this section, we show that with this simple choice of particle content, there exist parameter choices which will lead to CMB predictions identical to those of ${\cal R} + {\cal R}^2$ model of inflation of Starobinsky \cite{Starobinsky:1980te}.

\subsection{Slow roll inflation}

Irrespective of how complicated the inflaton potential is, if the potential slow roll parameters, defined by
\begin{eqnarray}
 \epsilon_V &\equiv& \frac{M_{\rm pl}^2}{2} \left(\frac{V'}{V} \right)^2 \; , \\
 \eta_V &\equiv& \frac{M_{\rm pl}^2 V''}{V} \; . 
\end{eqnarray}
are small as compared to unity at the time when the pivot scale $k_*$ crossed the Hubble radius during inflation, the primordial scalar and tensor power spectra are given by power functions of the wave-number.
Thus, in slow roll inflation, the primordial scalar and tensor power spectra are given by
\begin{eqnarray}
P_s(k) &=& A_s \left( \frac{k}{k_*} \right)^{n_s-1} \; , \\
P_t(k) &=& A_t \left( \frac{k}{k_*} \right)^{n_t} \; , 
\end{eqnarray}
where, the amplitude of the scalar power spectrum $A_s$, 
the tensor to scalar ratio,
and the scalar spectral index $n_s$ are given respectively by
\begin{eqnarray}
A_s &\approx& \frac{V}{24 \pi^2 M_{\rm pl}^4 \epsilon_V} \; , \\
n_s &\approx& 1+2\eta_V-6\epsilon_V\; , \\
r &\approx& 16 \epsilon_V \; .
\end{eqnarray}
Given these, other quantities such as the amplitude of the tensor power spectrum $A_t$, 
 and the tensor spectral index $n_t$ 
 \begin{eqnarray}
 A_t &\approx& \frac{2 V}{3 \pi^2 M_{\rm pl}^4} \; , \\
n_t &\approx& - 2 \epsilon_V \; .
 \end{eqnarray}
can easily be found from the relations $A_t = r A_s$ and $n_t = - \left( \frac{r}{8} \right)$.

Recall that if the pivot scale $k_*$ goes out of the Hubble radius during inflation at an epoch which was $N_*$ e-foldings from the end of inflation, then we expect, for GUT scale inflation, $N_*$ to be between $50$ and $60$ and in the following, we shall set $N_*$ to 60.
For the potential given in Eq (\ref{eq:pot}), for any choice of the parameters ${\cal N}, R,f$ and $Q$, one can numerically find $\phi_{\rm end}$, the value of inflaton field when inflation ends and then use 
\footnote{It is worth noting that we have replaced $\epsilon_H$ by $\epsilon_V(\phi)$  in order to get this relation.}
\begin{equation}
 N(\phi) = \int_{\phi_{\rm end}}^{\phi} \frac{d \phi}{M_{\rm Pl} \sqrt{2 \epsilon_V(\phi)}} \; ,
\end{equation}
and find $\phi_*$, the value of inflaton field when the pivot scale exited the Hubble radius. Finally, one can find the corresponding value of slow-roll parameters corresponding to $\phi_*$ and hence the scalar and tensor power spectra.

\subsection{Numerical results}

Lets us now see how the CMB predictions (i.e. $A_s, n_s$ and $r$) in this scenario change as we explore the parameter space of $\cal N$, $R$, $f$ and $Q$.
We would like to restrict our attention to the region of parameter space which offers slow-roll inflation and which
yields $n_s$ and $r$ that are most compatible with the Planck measurements i.e. 2015 Planck TT,TE,EE+lowP data \cite{Ade:2015lrj}. This data implies that $n_s = 0.9652\pm0.0047$ at $1\sigma$ C.L. while $r < 0.099$ (for $k_* = 0.002 {\rm Mpc}^{-1}$) and $A_s = 2.2065^{+0.0763}_{-0.0738} \times 10^{-9}$. 

Notice that in slow-roll inflation, $n_s$ and $r$ are completely determined by the slow-roll parameters and hence do not depend on any overall multiplicative factor in the potential.
The value of the any overall factor in the potential e.g. $R$ in Eq (\ref{eq:pot2}), can be adjusted to ensure that $A_s$ matches the observed value and is thus determined by $A_s$ and not $n_s$ and $r$. 

When ${\cal N} = 0$, one recovers the minimal version of extranatural inflation. Its predictions for the spectral index and the tensor to scalar ratio are identical to those of natural inflation and this model is mildly ($> 2 \sigma$) disfavoured by the Planck data \cite{Ade:2015lrj}.
If we restrict our attention to the case $Q=1$, then, no matter what value of $\cal N$ one works with, it is only $R$ which will be redefined. This will not change the slow-roll parameters and hence will not change the spectral index and the tensor to scalar ratio. Thus, the CMB predictions in this case won't be any better than those of natural inflation.
Similarly, the case $Q=0$ will only redefine $C$. Moreover, since cosine is an even function, for any given $\cal N$, the sign of $Q$ is unimportant. By numerically solving the underlying equations, one can also find that 
(a) for $f \leq 1.5$, the assumption $\eta_V \ll 1$ no longer remains valid and therefore one of the slow roll conditions gets violated. Since we wish to restrict our attention to slow-roll inflation, we choose to investigate the cases with $f >2$ in this work. Large values of $f$ yield values of $n_s$ and $r$ which are inconsistent with the most recent data and hence  $2 \leq f \leq 3$;
(b) for a fixed value of $f$, the slow roll conditions get violated if $Q \geq 2$ or $Q < 0.5$. So, we restrict ourselves to the range $0.5 \leq Q \leq 1.5$. 

\begin{figure}
  \includegraphics[width = .45\textwidth]{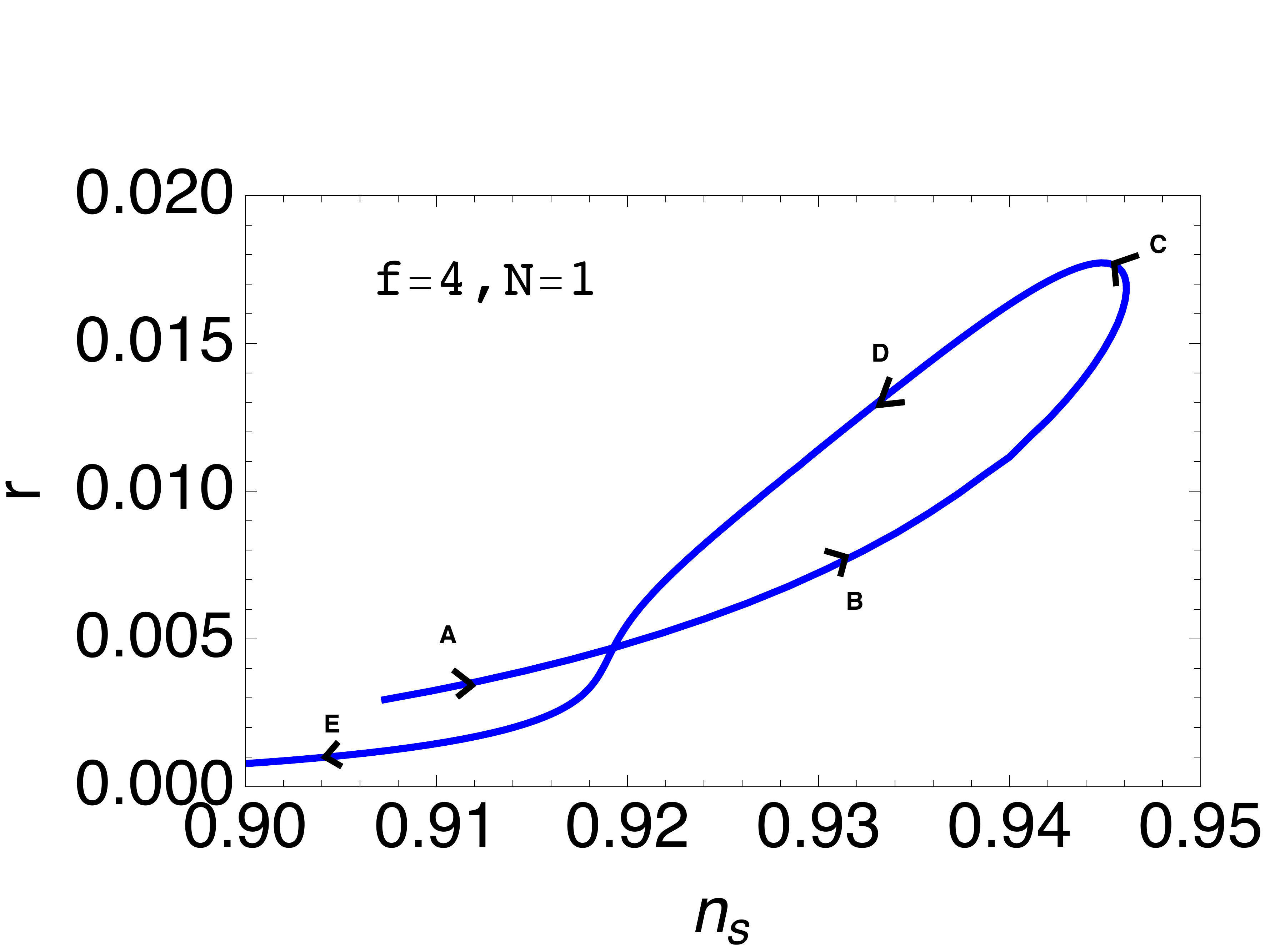}
  \caption
 {For a fixed $f = 4 M_{\rm Pl}$, changing the charge $Q$ on the additional fermion leads to a trajectory in $n_s$ and $r$ plane parameterized by $Q$ as shown here (for the case ${\cal N}=1$). As we increase the charge $Q$ from $0.5$ to $0.71$, we go from point {\bf A} to {\bf C} via point {\bf B}. At {\bf C}, there is a turning, the corresponding $Q=0.71$. Further increasing $Q$ from $0.71$ to $1.55$ takes us from {\bf C} to the ponts {\bf D} and {\bf E} along the path shown.}
  \label{zoomed}
\end{figure}

For a given combination of $\cal N$ and  $f$, as one changes $Q$, the charge of the additional fermion in the bulk, the predictions for $n_s$ and $r$ would change and we get trajectories in $n_s - r$ plane which are parameterized by $Q$.
Though the detailed shape of the curve depends on the choice of ${\cal N}$ and $f$, for any such choice, there is typically a range of $Q$ which will yield slow roll inflation and the corresponding trajectories in $n_s - r$ plane can then be found.
E.g., for the case ${\cal N} = 1$, $f = 4$, as one increases the charge from $Q_i = 0.5$ to $Q_f = 1.55$, one gets curves of the form shown in fig (\ref{zoomed}). 
For any given $f$, one can find the trajectories in $n_s$ and $r$ plane for the various values of $Q$ and one can then change $f$ and repeat this. Thus, for a given $\cal N$, one obtains a family of trajectories in $n_s - r$ plane. Let us now look at what happens as we choose various values of $\cal N$.

\subsubsection{${\cal N} = 1$}
\begin{figure}
  \includegraphics[width = .475\textwidth]{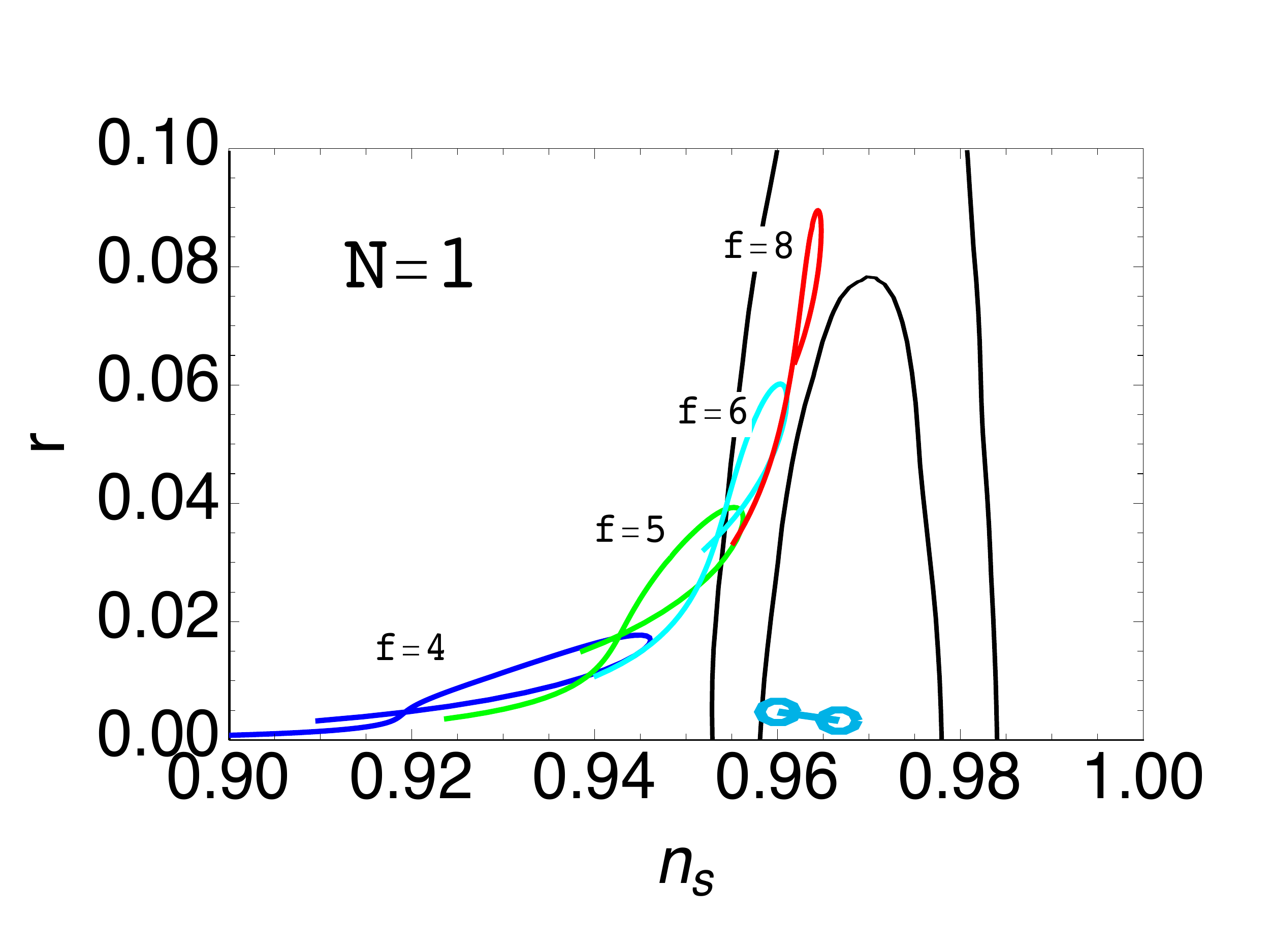}
  \caption
 {When ${\cal N} = 1$, no matter what value of $f$ and $Q$ are chosen, one never obtains the values of $n_s$ and $r$ which are inside the $1 \sigma$ contours of Planck TT,TE,EE+lowP data. Thus, there is thus no hope of obtaining CMB predictions similar to those of Starobinsky model.
}
  \label{N_1}
\end{figure}

The case ${\cal N} = 1$ corresponds to two fermions in the bulk. The charge of the first fermion has been set to $+1$ while that of the second one is Q. For this case, as shown in fig (\ref{N_1}), as one plots the family of trajectories corresponding to different $f$ and different $Q$, one finds that no choice of parameters brings us inside the $1 \sigma$ contours.

\subsubsection{${\cal N} = 2$}

We now argue that for the case with ${\cal N} = 2$, i.e. three fermions in the bulk, there exists a range of values of $f$ and $Q$ for which the CMB predictions improve significantly. 
E.g. if one chooses $R = 28.7$ (in units of reduced Planck length), $f=2.5$ and $Q=0.582$, then one finds (for $N_* = 60$) that, $n_s = 0.973$, $A_s = 2.19 \times 10^{-9}$, $r = 0.0036$. This must be compared with the predictions for Starobinsky model (for $N_* = 54$): $r = 0.004$ while $n_s = 0.963$.

\begin{figure}
  \includegraphics[width = .475\textwidth]{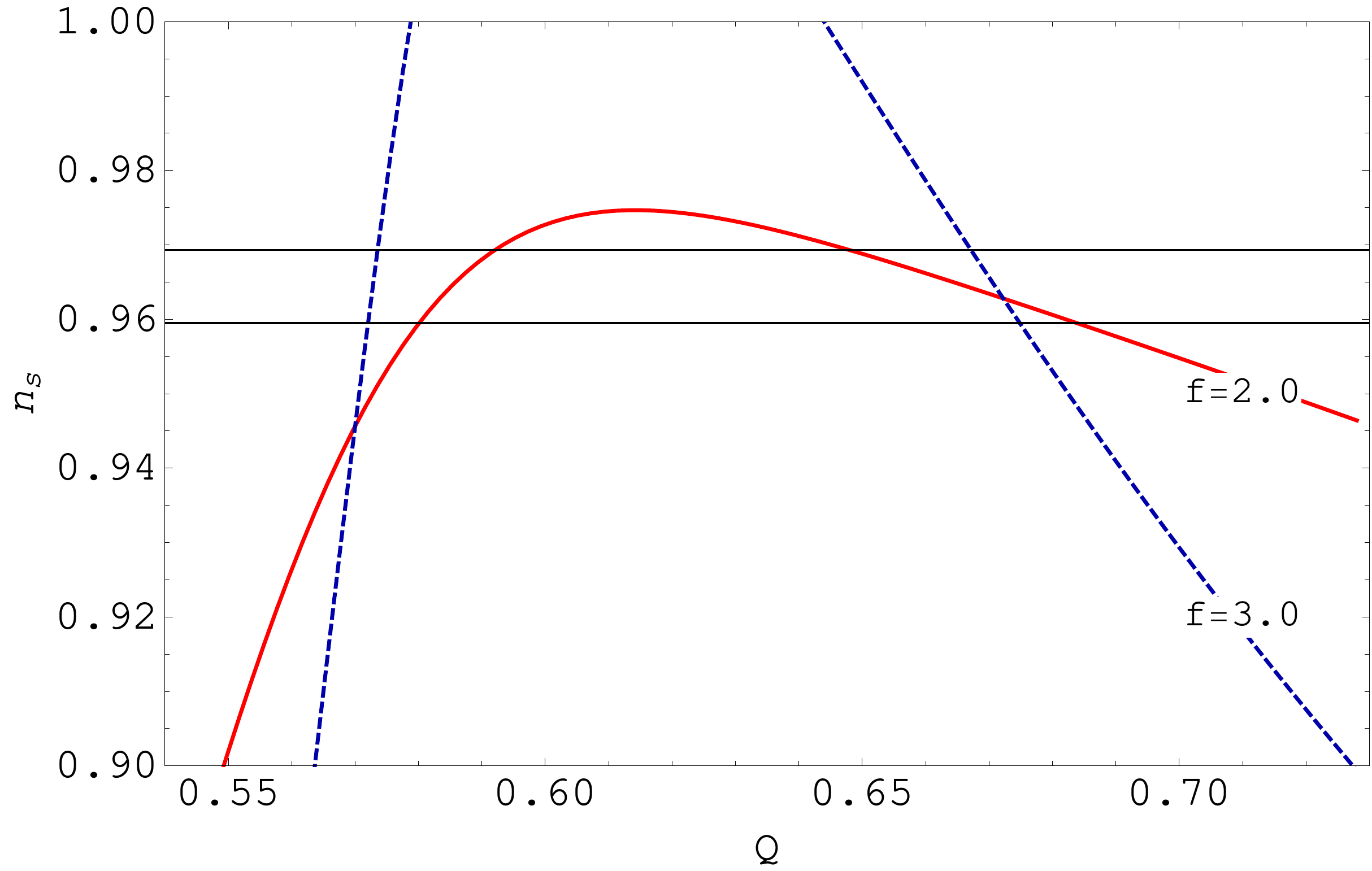}
  \caption
 {The behaviour of $n_s$ as one changes $Q$ for a fixed value of $f$ (in units of reduced Planck mass). The horizontal lines correspond to the $1 \sigma$ limits on $n_s$ for 2015 Planck TT,TE,EE+lowP data.
}
  \label{ns_Q}
\end{figure}

\begin{figure}
  \includegraphics[width = .475\textwidth]{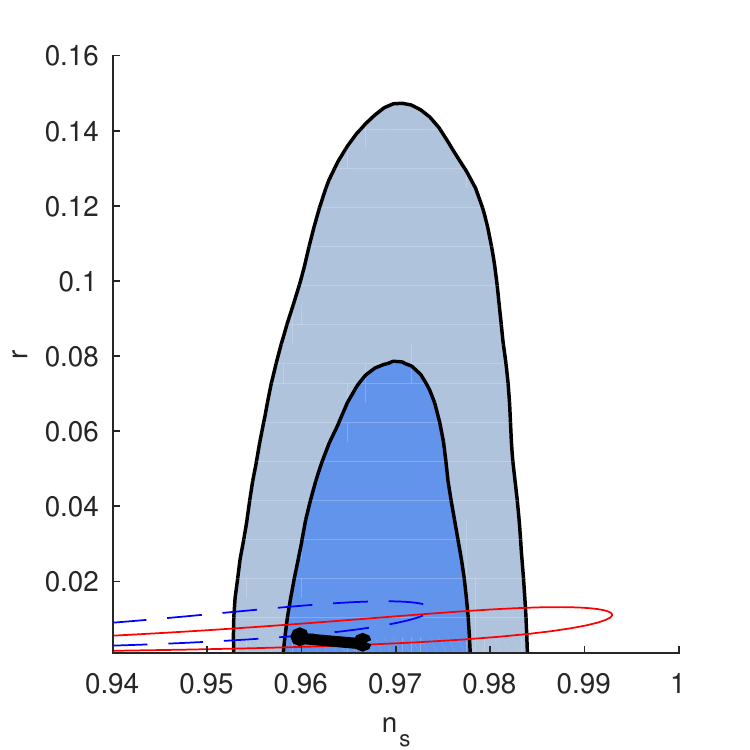}
  \caption
 {The trajectories in $n_s-r$ plane as one increases $Q$ for a fixed $f$. The two curves correspond to $f=3.0 M_{\rm Pl}$ (dashed) and $f=2.5 M_{\rm Pl}$ (solid) respectively. The predictions of Starobinsky model for $n_s$ and $r$ are also shown for reference. The shaded regions show the 1$\sigma$ and 2$\sigma$ contours for 2015 Planck TT,TE,EE+lowP data.
}
  \label{ns_r}
\end{figure}

\begin{figure}
  \includegraphics[width = .475\textwidth]{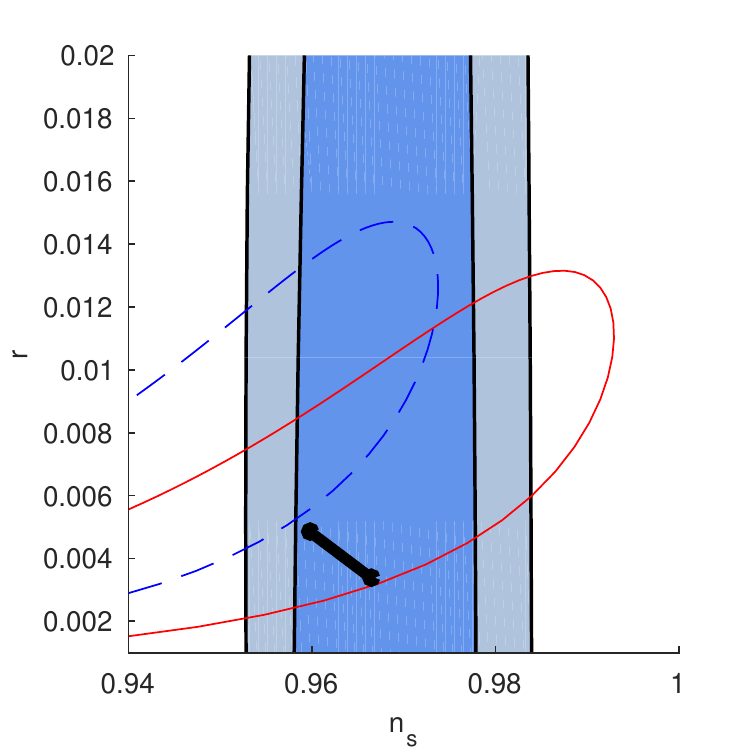}
  \caption
 {This is a ``zoomed-in" form of the previous figure: this shows that for $f$ between $2.5 M_{\rm Pl}$ (solid) and $3.0 M_{\rm Pl}$ (dashed) 
 and $Q$ between $0.55$ to $0.75$, there exists a combination of $f$ and $Q$ which will give predictions identical to those of Starobinsky model.
}
  \label{ns_r_Q_zoomed}
\end{figure}

If we restrict our attention to the range
\begin{equation} \label{eq:range}
 2.0 < f < 3, ~ 0.5 < Q < 0.75 \; ,
\end{equation}
firstly, as is shown in fig (\ref{ns_Q}), for each value of $f$ in the above range, there exist values of $Q$ for which $n_s$ lies within the $1\sigma$ allowed region.
When $f$ is in this range, increasing $Q$ from 0.5 first increases $n_s$ such that for a small range of values of $Q$, 
$n_s$ does fall within the observationally preferred range. It then reaches its local maximum value and then decreases. While decreasing, the value of $n_s$ again falls within the observationally preferred range.
Secondly, as fig (\ref{ns_r}) and fig (\ref{ns_r_Q_zoomed}) indicate, for this range of parameters, the tensor to scalar ratio stays below $0.02$ and this is true irrespective of the value of $Q$ and $f$. 

In summary, when one is in the range specified by Eq (\ref{eq:range}), $r$ stays below 0.02; changing $Q$ essentially changes $n_s$, $f$ has a small effect on $n_s$ and $r$, while, as was mentioned earlier, $R$ essentially determines $A_s$. 
One finds that there must be some combination of $f$ and $Q$ which leads to $n_s$ and $r$ which are identical to those obtained in Starobinsky model (see e.g. fig (\ref{ns_r_Q_zoomed})).

We thus learn that adding fermions in the bulk with appropriately chosen charges can improve the CMB predictions of extranatural inflation. 
Moreover, CMB data suggests that ${\cal N} = 2$, $R \approx 29 ~ M_{\rm Pl}^{-1}$, $f \approx 2.5 M_{\rm Pl}$ and $Q \approx 0.58$.


\subsection{Constraints on derived parameters} \label{sec:scales}

Since the basic parameters of this scenario are constrained by CMB data, it may be a good idea to find the constraints on the other, derived parameters.
But before we do so, since there are many scales in the problem, it will be a good idea to get some feel for their relative hierarchies before proceeding. 
Let $\ell_P$ be the 4D Planck length, $L$ ($=2 \pi R$) be the size of the extra dimension, $M_p$ be the 4D Planck mass and $M_p^{(5)}$ be the 5D Planck mass, then \cite{Zwiebach:2004tj}, 
\begin{equation} \label{eq:qgscales}
L = \left( \frac{M_p}{M_p^{(5)}} \right)^3 \ell_P \; ,
\end{equation}
and this implies that, 
\begin{equation} \label{eq:qgscales-2}
\frac{L^{-1}}{M_p^{(5)}} = \frac{L^{-1}}{M_p} \frac{M_p}{M_p^{(5)}} = \left( \frac{M_p^{(5)}}{M_p} \right)^2 \; .
\end{equation}
Now in 5D, the energy scale of quantum gravity is $M_P^{(5)}$ while $L^{-1}$ is the cut-off scale of 4D EFT. So, 
$L^{-1}$ must be smaller than $M_p^{(5)}$ so, from Eq (\ref{eq:qgscales}) and Eq (\ref{eq:qgscales-2}), 
\begin{equation}
M_p > M_p^{(5)} \; .
\end{equation}

Using Eq (\ref{eq:decayconst}) and the fact that we typically need $f > M_p$ to be consistent with data, one concludes that 
\begin{equation}
g_4 \left( \frac{M_p}{M_p^{(5)}} \right)^3 < 1.
\end{equation}
Since 5D Planck mass (i.e. $M_p^{(5)}$) is smaller than 4D Planck mass (i.e. $M_p$), the above inequality implies that we must have
\begin{equation}
g_4 \ll 1 \; .
\end{equation}
Moreover, since the 5D gauge coupling has a negative mass dimension, Eq (\ref{eq:unitarity}) implies that the unitarity bound of the theory is $E_{\rm strong} = 1/(2 \pi R g_4^2)$, so that when the energy scale of any process is of this order, perturbative unitarity gets violated. Thus, 
\begin{equation}
\frac{E_{\rm strong}}{f} = \frac{1}{g_4} \gg 1 \; .
\end{equation}

Similarly, Eq (\ref{eq:pot}) implies that $V \sim L^{-4}$ while Friedman equation implies that the Hubble parameter during inflation is given by $H^2 = \frac{8 \pi V}{3 M_p^2}$
so that,
\begin{equation}
  H = \sqrt{\frac{8\pi}{3}} \frac{1}{L^2 M_p} \; ,
\end{equation}
 and hence
\begin{equation}
\frac{H}{L^{-1}} = \sqrt{\frac{8\pi}{3}} \frac{L^{-1}}{M_p} \ll 1 \; .
\end{equation}
The above analysis implies that 
\begin{equation}
E_{\rm strong} \gg f \gg M_p > M_p^{(5)} > L^{-1} \gg H \; .
\end{equation}

Let us now find the numerical values of all of these ratios, given the best fit values of $R$ and $f$. 
Obviously, $M_p = \sqrt{8 \pi} M_{\rm Pl} = 5.013 M_{\rm Pl}$ and 4D gauge coupling $g_4 = (2\pi R f)^{-1}$ is given by 
\begin{equation}
g_4 = 0.0022 \; .
\end{equation}
The energy scale $E_{\rm strong} = 1.13 \times 10^3 ~M_{\rm Pl}$, the five dimensional Planck scale is given by 
$M_p^{(5)} = 0.518 ~M_{\rm Pl}$, the scale $R^{-1} = 0.035 ~M_{\rm Pl}$ and, finally, $H = 1.77 \times 10^{-5} ~M_{\rm Pl}$.

Any scenario which UV completes extranatural inflation then needs a mechanism to keep $g_4$ at $2.2 \times 10^{-3}$ and $R \approx 30 M_{\rm Pl}^{-1}$. Adding bulk fermions will surely destabilize the potential of the radion and cause the extra dimension to contract to Planckian values \cite{Appelquist:1982zs}. 
In order to keep the size of the extra dimension fixed to the desired value, the vev of the radius modulus must stay put at the desired value.
There are many ways of dealing with this problem and the most harmless one is to use stabilizer fields a la Goldberger-Wise \cite{Goldberger:1999uk} to ensure that the extra dimension stays sufficiently large. 
Since these fields need not be charged under the gauge group of the bulk gauge field, the inflaton potential and hence inflationary predictions will not be affected by employing this mechanism. 
This mechanism will work only if the radion is not excited during inflation. The condition for this is that the mass of radion must be large as compared to $H$. 

\section{Summary and discussion} \label{discussion}

There seems little doubt that cosmological experiments of the next generation will be able to reduce the uncertainties in various inflationary observables to a very high degree. E.g. it is expected that within a decade, the uncertainties in the tensor to scalar ratio i.e. $\sigma(r)$ will be of the order of $0.0005$ \cite{Abazajian:2016yjj} while those in the running of the scalar spectral index ($\alpha_s$) shall be of the order of $0.0025$ \cite{Munoz:2016owz}. 
It thus seems that in near future, we shall uncover the shape of the inflaton potential with unprecedented accuracy.
Given this optimistic state of affairs, one must ensure that the mechanisms which lead to the inflaton potential are completely trustworthy from theoretical perspective.

The CMB observations of even the current generation have imposed tight observational constraints on many scenarios of cosmic inflation. 
A model whose CMB predictions are compatible with the observations of Planck experiment is ${\cal R} + {\cal R}^2$ model of inflation of Starobinsky \cite{Starobinsky:1980te} (see also \cite{Ellis:2013xoa}). 
If one considers higher dimensional corrections to Einstein-Hilbert action i.e.
$\frac{\cal R}{2\kappa^2} + \alpha {\cal R}^2 + \beta {\cal R}_{\mu \nu} {\cal R}^{\mu \nu} 
+ \gamma {\cal R}_{\mu \nu \rho \sigma} {\cal R}^{\mu \nu \rho \sigma} $, 
one can use Chern-Gauss-Bonnet Theorem  
(i.e. the fact that ${\cal R}^2 - 4 {\cal R}_{\mu \nu} {\cal R}^{\mu \nu} + {\cal R}_{\mu \nu \rho \sigma} {\cal R}^{\mu \nu \rho \sigma}$ is a topological invariant) to eliminate the term with Riemann tensor.
Now, Starobinsky model is based on the assumption that among $\kappa$, $\alpha$, $\beta$ 
in 
\begin{equation}
S = \int d^4 x \sqrt{-g} \left(
\frac{\cal R}{2\kappa^2} + \alpha {\cal R}^2 + \beta {\cal R}_{\mu \nu} {\cal R}^{\mu \nu} 
\right) 
+ \cdots \; ,
\end{equation}
one can pick only the $\frac{\cal R}{2\kappa^2} + \alpha {\cal R}^2$ terms and adjust the coefficients to ensure that one obtains CMB predictions. 
This is sensible because, among one-loop quantum corrections to the Einstein-Hilbert action, the terms 
other than $\frac{\cal R}{2\kappa^2} + \alpha {\cal R}^2$ vanish when the metric is conformally flat (e.g., a spatially flat FRW metric) as happens after sufficient duration of inflation.
However, if $\kappa$ is 4D reduced Planck length, CMB observations imply that the dimensionless coefficient $\alpha$ is ${\cal O}(10^9)$.
Given this, one might wonder whether assuming $\alpha$ to be too large and $\beta$ to be negligible may appear to involve some fine-tuning but notice that since $\beta$ is small due to a symmetry reason, its smallness may be protected from loop corrections.
Similarly, if one resorts to coupling the inflaton field non-minimally to gravity in order to obtain successful CMB predictions, super-Planckian vevs may make it hard to justify including just the one higher dimensional operator ${\cal R} \phi^2$ while ignoring all others (see however, the discussion in section 5.1 of \cite{Baumann:2010ys}). 
But, the super-Planckian field excursion in the Einstein frame field in both the cases discussed (Starobinsky as well as non-minimal coupling) suggests that one may need to include even higher order terms.
Given this, a scenario of large field inflation which does not suffer from such uncertainties and nonetheless makes successful predictions which can be tested in next generation experiments must be seriously sought. In this work, an attempt has been made to find one such model.

Ultra-violet sensitivity of large field inflation persuades one to resort to symmetries to protect the inflaton potential. The minimal natural inflation uses global symmetries to deal with this but since there can be no global symmetries in quantum gravity \cite{Kallosh:1995hi,Banks:2010zn}, a better alternative is to employ gauge symmetries for this task. 
Extranatural inflation does exactly this: the inflaton is the zero mode of the fifth component of a bulk Abelian gauge field whose potential is generated by charged fermions present in the bulk.
The required value of gauge coupling to achieve the same however turns out to be quite small. This small value of this 4D gauge coupling has been a topic of discussion in the literature since the failure to obtain such small values of the gauge coupling in known UV completions such as string theory \cite{Banks:2003sx} has inspired the famous Weak Gravity Conjecture \cite{ArkaniHamed:2006dz}.
The exact value of gauge coupling is thus of paramount importance.

However, the most minimal version of extra-natural inflation gives predictions identical to those of natural inflation which is mildly disfavoured by current CMB data. 
One may wonder whether one could have a variation of the minimal version of extranatural inflation which fits the data and then it is for such a model the value of 4D gauge coupling $g_4$ is to be found. 
In this work, we studied the effect of additional charged, light, fermions in the bulk on the CMB predictions of extranatural inflation.
We have found that the one needs to add at least two more fermion species in the bulk in order to improve the fit to CMB data and if the radius of the extra dimensions is $R \approx 29 M_{\rm Pl}^{-1}$, $f = 2.5 M_{\rm Pl}$ and the charges of the additional fermions are $Q=0.58$ (in units of charge of the fermion which generates the potential of extranatural inflation), one obtains CMB predictions very close to those of Starobinsky model. One can readily determine the corresponding value of $g_4$ and it turns out to be $g_4 = 0.0022$. 
Since this value still turns out to be too small, we have essentially shown that though adding more fermions can help improve the fit to CMB, this does not resolve the problem of smallness of gauge coupling. 
In any case, it is this value of the 4D gauge coupling, which needs to be targeted in the UV completions of extranatural inflation.

In summary, we have presented a model of cosmic inflation which has the merit that its predictions can be identical to those of Starobinsky model and which can potentially be free from all issues of UV sensitivity provided one can find a UV completion in which the 4D gauge coupling turns out to be as small as required. 
Apart from looking for appropriate UV completions, in future one could make use of the light bulk fermions in the model to tackle with other problems of the physics of the early universe e.g. matter-anti-matter asymmetry or dark matter. Since the masses and charges of such fermions are already constrained by CMB data, this can be a very interesting exercise.

 \noindent{\bf Acknowledgements:}
    G.G. would like to thank Koushik Dutta (Saha Institute of Nuclear Physics, Kolkata) and Tirtha Sankar Ray (Indian Institute of Technology, Kharagpur) for discussions. G.G. would also like to thank Gary Shiu (University of Wisconsin, Madison) for clarifying several issues related to extranatural inflation. J.P. would like to thank the Navajbai Ratan Tata Trust (NRTT) for financial support.

\end{document}